\documentclass[preprint2]{aastex}
\usepackage{color}

\shorttitle{Solar Microwave Zebra Pattern Burst}
\shortauthors{B.L. Tan}

\begin{document}

\title{A Very Small and Super Strong Zebra Pattern Burst at the Beginning of a Solar Flare}

\author{Baolin Tan$^1$, Chengming Tan$^1$, Yin Zhang$^1$, Jing Huang$^1$, Hana M\'esz\'arosov\'a$^2$, Marian Karlick\'y$^2$, Yihua Yan$^1$}
\affil{$^1$ Key Laboratory of Solar Activity, National
Astronomical Observatories of Chinese Academy of Sciences, Beijing
100012, China, Email: bltan@nao.cas.cn} \affil{$^2$ Astronomical
Institute of the Academy of Sciences of the Czech Republic,
Ond\v{r}ejov 15165, Czech Republic}

\begin{abstract}

Microwave emission with spectral zebra pattern structures (ZPs) is
observed frequently in solar flares and the Crab pulsar. The
previous observations show that ZP is only a structure overlapped
on the underlying broadband continuum with slight increments and
decrements. This work reports an extremely unusual strong ZP burst
occurring just at the beginning of a solar flare observed
simultaneously by two radio telescopes located in China and Czech
Republic and by the extreme ultraviolet (EUV) telescope on board
NASA's satellite Solar Dynamics Observatory on 2013 April 11. It
is a very short and super strong explosion whose intensity exceeds
several times that of the underlying flaring broadband continuum
emission, lasting for just 18 s. EUV images show that the flare
starts from several small flare bursting points (FBPs). There is a
sudden EUV flash with extra enhancement in one of these FBPs
during the ZP burst. Analysis indicates that the ZP burst
accompanying EUV flash is an unusual explosion revealing a strong
coherent process with rapid particle acceleration, violent energy
release, and fast plasma heating simultaneously in a small region
with short duration just at the beginning of the flare.

\end{abstract}


\section{Introduction}

Microwave bursts are frequently observed from the Sun (Elgaroy
1959; Tanaka \& Enome 1970; Dulk 1985; Bastian, Benz, \& Gary
1998), flare stars (Bastian et al. 1990), neutron stars
(McLaughlin et al. 2006), and even the Crab pulsar (Hankins \&
Eilek 2007). Spectral fine structures in microwave bursts are
believed to contain abundant information of particle acceleration,
energy release, and other non-thermal processes in magnetized
astrophysical plasmas. In particular the microwave zebra pattern
(ZP), a kind of fine structure superposed on the solar radio
broadband continuum spectrogram, which consists of several almost
parallel and equidistant stripes. It is an interesting and
intriguing phenomenon, which may reveal the original information
of the flaring source region, such as the magnetic field and its
configurations, particle acceleration, and plasma instabilities in
the source region where the energy release takes place (Kuijpers
1975; Chernov et al. 2005; Tan et al. 2012; Chernov 2006; Hankins
\& Eilek 2007; Karlick\'y 2013). So far, the nature and formation
mechanism of microwave ZP structures is still a controversial
problem which has been discussed widely for more than 40 years
(Rosenberg \& Tarnstrom 1972; Zheleznyakov \& Zlotnik 1975;
Fomichev \& Fainshtein 1981; LaBelle et al. 2003; Tan 2010; Chen
et al. 2011). According to a recent statistical classification,
microwave ZPs can be sorted into three types, and different type
will reveal different physical non-thermal processes by different
mechanism (Tan et al. 2014).

In the previous observations, the microwave ZP is always only a
structure overlapped on the underlying broadband continuum with
slight increments and decrements. This work reports an extremely
unusual ZP burst just at the beginning of a solar flare observed
simultaneously by two radio telescopes located in China and Czech
Republic and by the extreme ultraviolet (EUV) telescope on board
NASA's satellite Solar Dynamics Observatory (SDO). It is a very
short and super strong explosion which may reveal a rapid particle
acceleration and fast plasma heating in a small region and in a
short time. Section 2 presents the observations and physical
analysis, and Section 3 is the conclusion and and some
discussions.

\section{Observations and Physical Analysis}

\subsection{Observation data}

In this work, observation data are obtained from the following
instruments:

(1) The Chinese Solar Broadband Radio Spectrometers at Huairou
(SBRS/Huairou)

SBRS is an advanced solar radio telescope with broad frequency
bandwidth, and super high temporal- and spectral-resolutions,
which can distinguish super fine structures from the spectrogram
(Fu et al. 1995, 2004; Yan et al. 2002). Its daily observational
window is 0:00-8:00 UT during winter seasons and 23:00-9:00 UT
during summer seasons. It includes three parts: 1.10 - 2.06 GHz
(with antenna diameter of 7.0 m), 2.60 - 3.80 GHz (with antenna
diameter of 3.2 m), and 5.20 - 7.60 GHz (share the same antenna of
the second part). The antenna points to the solar disk center
automatically controlled by a computer. The spectrometer receives
the total flux of solar radio emission with dual circular
polarization (left- and right-handed circular polarization, LCP
and RCP), and the dynamic range is 10 dB above the quiet solar
background emission. The observation sensitivity is:
$S/S_{\bigodot}\leq 2\%$, here $S_{\bigodot}$ is the quiet solar
background emission. In this work we use the observation data at
frequency of 2.60 - 3.80 GHz with cadence of 8 ms and frequency
resolution of 10 MHz.

(2) Ond\v{r}ejov radiospectrograph in the Czech Republic
(ORSC/Ond\v{r}ejov)

ORSC is a broadband spectrometer located at Ond\v{r}ejov, the
Czech Republic. It receives solar radio total flux at frequencies
of 0.80 - 5.00 GHz during 2000 - 2013 (Ji\v{r}i\v{c}ka et al.
1993). Its daily observational window is 7:00 - 16:00 UT in winter
seasons and 6:00 - 17:00 UT in summer seasons. In this work we use
the observation data at frequency of 2.00 - 5.00 GHz with a
cadence of 10 ms and a frequency resolution of 12 MHz.

SBRS/Huairou and ORSC/Ond\v{r}ejov have an overlapping
observational window 7:00 - 8:00 UT during winter seasons and
6:00-9:00 UT during summer seasons. This common window provides a
good opportunity to observe some solar eruptions simultaneously.

(3) Atmospheric Imaging Assembly (AIA) on board NASA's satellite
Solar Dynamics Observatory (SDO)

AIA obtains the full-disk snapshot images of the solar corona and
transition region with a cadence of 12 s and pixel size of 0.6$''$
at UV and EUV wavelengths centered on specific lines: 1700 \AA,
1600 \AA, 335 \AA, 304 \AA, 211 \AA, 193 \AA, 171 \AA, 131 \AA,
and 94 \AA. They are formed at different temperatures in the solar
chromospheric and coronal plasmas (Lemen et al. 2012), and can
present the detailed configurations and fast variations of solar
chromosphere and corona.

(4) Soft X-ray (SXR) telescope on Geostationary Operational
Environment Satellites (GOES)

The GOES provides continuous monitoring of integrated full-disk
solar SXR intensity at 0.5-4 \AA~ channel and 1-8 \AA~ channel
with cadence of 3 s. It is the most important indicator to show
solar flare processes.

\subsection{Main Results}

\subsubsection{Microwave Spectral Observation}

The left panel of Fig.1 presents the microwave spectrograms
observed by SBRS/Huairou at LCP and RCP. It shows that there is a
bright patch at frequency of 2.65 - 3.10 GHz during 06:58:26 -
06:58:44 UT, just at the very beginning of an M6.5 class flare.
This panel is an expanding one showing that the bright patch is
similar to a ZP structure which lasts for about 18 s, has 4 curved
stripes. The central frequency is 2.87 GHz. The frequency
separation of adjacent stripes is about 90 MHz, approximately a
constant. The frequency bandwidth of the whole structure is 450
MHz, and the relative bandwidth is 15\%. The comparison between
RCP and LCP indicates that the bright patch burst is moderate LCP
with polarization degree of about 35\%. However, we have no enough
confidence to confirm it as a ZP structure because there are
several bad frequency channels on the spectrogram observed by
SBRS/Huairou.

\begin{figure*}[ht]  
\begin{center}
   \includegraphics[width=16.4 cm]{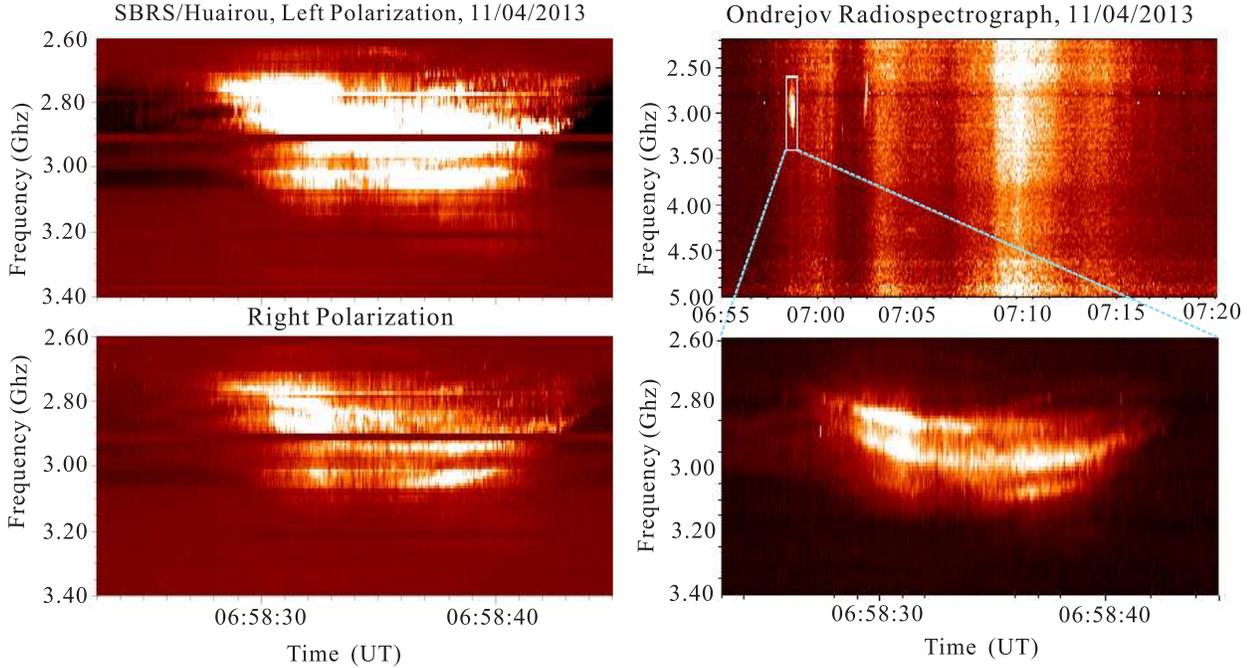}
\caption{Spectrograms of the microwave zebra pattern burst at the
beginning of an M6.5 flare on 2013 April 11. The left panels are
obtained by the Chinese Solar Broadband Radio Spectrometers at
Huairou at left- and right-handed circular polarization. Right
panel is obtained by the Ond\v{r}ejov radiospectrograph in the
Czech Republic.} \label{fig:source}
\end{center}
\end{figure*}

It is very lucky that ORSC/Ond\v{r}ejov observed also the same
microwave burst without any influence of bad frequency channels.
The right-upper panel of Fig.1 is the whole spectrogram at
frequency of 2.00 - 5.00 GHz during 06:55 - 07:20 UT. Here, we
find a broadband microwave continuum burst occurring from the
beginning of the flare and lasting to after the flare maximum in
the whole frequency range of the spectrometer. The most
highlighted is that there is a small and extremely bright patch
overlapped on the broadband continuum spectrogram which is marked
in a white box. The enlarged spectrogram of the bright patch is
presented in the right-bottom panel of Fig. 1. It is much clearer
than the spectrogram observed by SBRS/Huairou. Here, we find again
that the bright patch consists of four crescent-shaped stripes,
and the frequency range, stripe separations, brightness, duration,
and the shapes of each stripes are strictly identical to the
spectrogram obtained by SBRS/Huairou. These properties show that
this spectral structure is most coinciding with a ZP phenomenon,
different from the other stripe-like spectral events, such as
fiber bursts, lace bursts, and fundamental-harmonic structures
(Huang \& Tan 2012) etc. Fiber bursts are always with constant
frequency drifting rates and random frequency separations (Bernold
\& Treumann 1983, Wang \& Zhong 2006). Lace burst has only one or
two stripes with rapid variations of frequency drifting rates
(Karlick\'y et al. 2001). The fundamental-harmonic structures
generally has only two or three stripes and frequencies are in
harmonic ratio (Stahli, Magun, \& Schanda, 1987). Additionally,
scrutinizing analysis shows that each stripe is composed of many
millisecond spike bursts. It is a complicated structure of ZP and
spike bursts. This is the first time that an identical microwave
ZP structure observed simultaneously by two separate telescopes
away from more than seven thousand kilometers. Here, once again,
it shows that the zebra event is an extremely strong burst which
is much brighter than the basal flaring continuum radiation.

\begin{figure*}[ht]  
\begin{center}
   \includegraphics[width=11 cm]{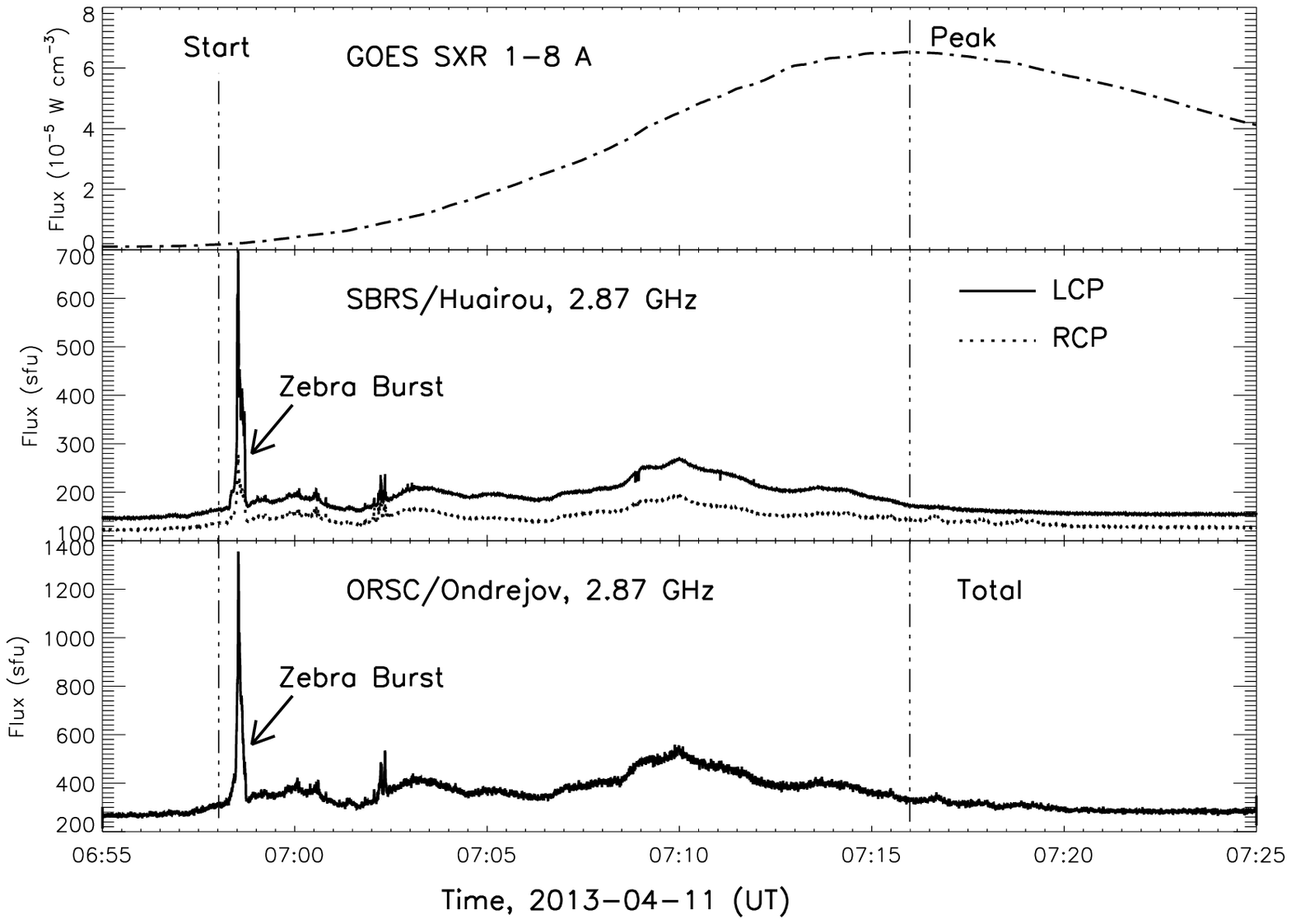}
\caption{Profiles of microwave emission intensities at 2.87 GHz at
left- and right-handed circular polarization (LCP and RCP)
observed by SBRS/Huairou and by ORSC/Ond\v{r}ejov (total), and the
GOES soft X-ray intensities at 1 - 8 \AA (SXR) in the M6.5 flare
on 2013 April 11. It shows that the zebra burst is so strong that
it exceeds several times of the basal flaring continuum emission.}
\label{fig:source}
\end{center}
\end{figure*}

The previous observations show that stripes of microwave ZPs
always have only a little emission enhancements and decrements
superposed on the basal flaring broadband continuum (see the
review of Chernov 2006). Different from the previous observations,
the most unusual property of the ZP event here is that it is a
super strong and isolated burst. Fig.2 presents the profiles of
microwave emission fluxes at the central frequency (2.87 GHz) of
the ZP structure at LCP and RCP obtained by SBRS/Huairou, the
total flux obtained by ORSC/Ond\v{r}ejov, and SXR at 1-8 \AA~
obtained by GOES during the flare. SXR observations show that the
flare takes place in active region NOAA 11719 located very close
to the solar disk center. It is a two-ribbon M6.5 class flare,
starts at 06:58 UT, reaches to maximum at 07:16 UT, and ends at
07:26 UT. The ZP event occurs just at the beginning of the flare.
Its maximum emission flux is so strong that exceeds several times
of the basal flaring continuum intensity. The total flux reaches
to 1350 sfu (sfu is the solar radio flux unit, 1 sfu $=10^{-22}
Wm^{-2}Hz^{-1}$), while the flaring continuum emission flux before
and after the ZP burst is only about 200 sfu, the net intensity of
the ZP event is about 1150 sfu. From the bandwidth and intensity,
the emission brightness temperature can be estimated as
$2.10\times10^{11}$ K. It is so strong that only coherent
mechanism can produce it. Therefore, we call such explosion as ZP
burst.

As a comparison, there are several microwave type III bursts
occurring about 2 minutes after the ZP burst, but their
intensities are much weaker than the ZP burst, see the bottom
panel of Fig. 4.

\subsubsection{EUV Imaging Observation}

Although we know a super strong microwave ZP burst taking place at
the beginning of the flare, but we do not know the location of the
source region and its configuration for lack of imaging
observations at the corresponding frequencies up to now. Similar
to the solar microwave emission, EUV emissions also have source
regions locating from chromosphere to the lower corona, it is
possible to get some indirect information of the source region of
the ZP burst from scrutinizing of the EUV snapshot imaging
observations obtained by AIA/SDO in the same duration.

The snapshot imaging observations show that the flaring region
brightens up continuously from the flare onset (06:58 UT) to its
maximum (07:16 UT) at UV and EUV wavelengths. There are two
reversed S-shaped ribbons during the flaring process. The
scrutinizing of the snapshot images reveals that the flaring
ribbons are consisting of several separated small bright points
before the flare onset at almost all UV and EUV wavelengths. The
white arrows in left-upper panel of Fig.3 indicate such bright
points occurring before the flare onset. At the initial phase,
bright points are very small and separated from each other. Then
they become larger and larger gradually. Finally, they connected
with each other and form two bright reversed S-shape ribbons. We
call such bright points as flare bursting points (FBPs).

\begin{figure*}[ht]  
\begin{center}
   \includegraphics[width=15 cm]{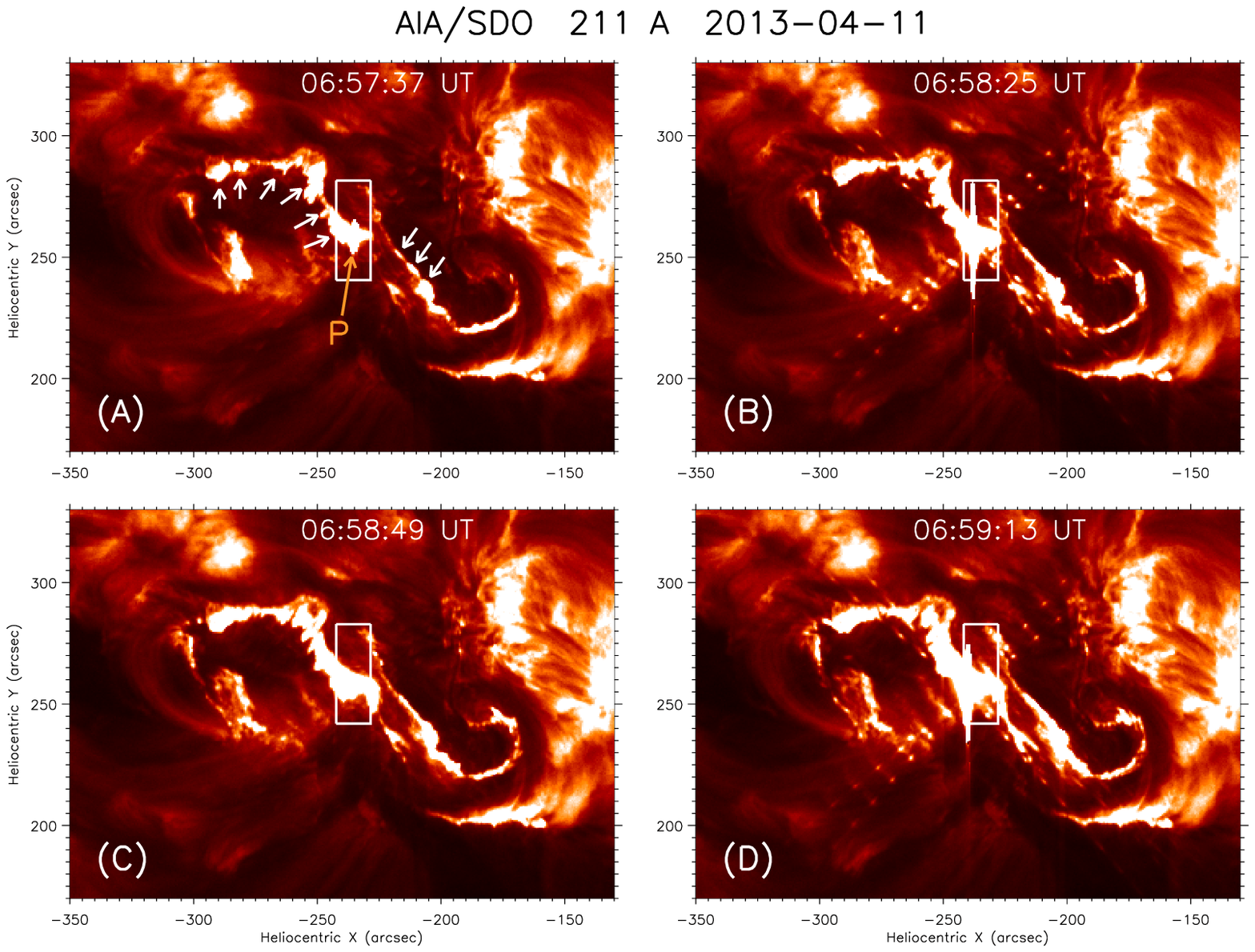}
\caption{EUV images at 211 \AA~ during the ZP burst obtained by
AIA/SDO on 2013 April 11. White arrows in panel (A) show EUV
bright points before the flare onset. They gradually connect with
each other and form the bright flare ribbons after the flare onset
showing in panel (B), (C), and (D). Among these bright points,
there is a sudden EUV flash during the ZP burst in the white box
marked in a yellow arrow (P).} \label{fig:source}
\end{center}
\end{figure*}

Among these FBPs, it is most interesting that there is one which
has a sudden strong extra enhancement at EUV wavelengths
overlapped on the background continuously brightening just during
the above microwave ZP burst. This FBP locates at the middle of
one ribbon centered in the flaring region (marked $P$ and a long
yellow arrow in the white box). Fig.3 presents images observed at
211 \AA~ at four different moments. Before the ZP burst it is just
a small bright point (left-upper panel, at 06:57:37 UT). During
the ZP burst it becomes so strong that the emission exceeds the
saturation level (right-upper panel, at 06:58:25 UT). At the end
of the ZP burst its brightness decreases slightly but the area
enlarges obviously (left-bottom panel, 06:58:49 UT). After the ZP
burst, it continues to brighten up along with the evolution of the
flare process.

Besides images of 211 \AA, the similar extra enhancements are also
occurring at wavelengths of 171 \AA, 193 \AA~ and 304 \AA~ in the
same region with saturations. However, different EUV wavelengths
will saturate at different levels for their different emission
strengths. The images at 94 \AA, 131 \AA~ and 335 \AA~ have a
moderate extra enhancement with respect to its background
continuous brightening although they do not exceed the saturation
level in the above small region. We call all of these small sudden
strong enhancements at EUV wavelengths as EUV flash. The width of
the EUV flash region is less than 3$''$ ($<$ 2200 km) and the
length is less than 7$''$ ($<$ 5000 km). It lasts for $<$ 24 s
which is very close to the duration of the above ZP burst.

\begin{figure*}[ht]  
\begin{center}
   \includegraphics[width=12 cm]{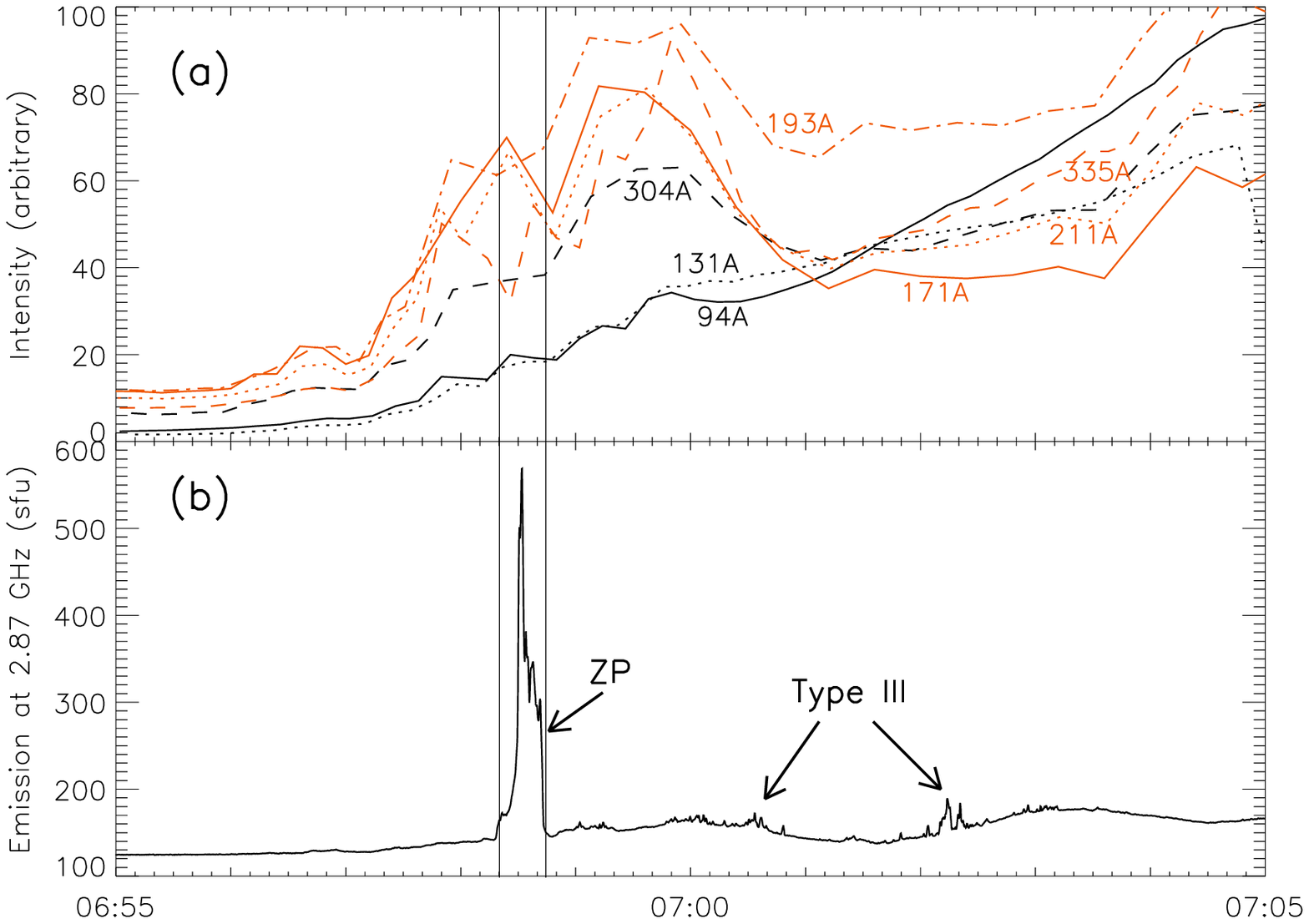}
\caption{(a) is profiles of the integral EUV brightness in the
white box at wavelengths of 94 \AA, 131 \AA, 171 \AA, 193 \AA, 211
\AA, 304 \AA, and 335 \AA. (b) is the profile of microwave
emission at frequency of 2.87 GHz centered at the ZP burst}
\label{fig:source}
\end{center}
\end{figure*}

In order to show the emission variations at different wavelengths
in the above small region, Fig.4 plots profiles of the integral
emission intensity calculating at different EUV wavelengths in a
rectangle box surrounding the above EUV flash FBP (white box in
Fig. 3) during 06:55 - 07:05 UT. As a comparison, the profile of
microwave emission at the central frequency of the ZP structure
(2.87 GHz) is also overplotted in Fig.4. Here we find that all
profiles increase continuously along with the flaring process. It
is reasonable because the small EUV flash region is just located
near the center of the flaring region, and the period is just at
the flare initial rising phase. The most important is that there
are some extra enhancements on the profiles at wavelengths of 171
\AA, 193 \AA, 211 \AA~ and 335 \AA~ during the ZP burst which
exceed the background gradual increments (upper panel of Fig. 4).
Here, the extra enhancement at 94 \AA, 131 \AA~ and 304 \AA~ is
not so obvious as above wavelengths. Of course, in above
calculation, the rectangle box is slightly larger than the EUV
flash region, the integration may smooth the variations of EUV
emission to some extent. The cadence is only 12 s (24 s
sometimes), the above results only reflect approximately the
variations during the ZP burst at different EUV wavelengths.

\subsection{Physical Analysis}

The EUV flash indicates that the emission has a sharp extra
enhancement in a small region during the microwave ZP burst. It is
well-known that different EUV emission lines may have different
formation temperatures produced in solar chromosphere and corona
(Lemen et al. 2012). The EUV emission intensity is proportional to
the square of plasma density and the gradient of temperature. It
is only the short rapid heating by energetic electrons that can
produce a sharp enhancement of EUV emissions (Tandberg-Hanssen \&
Emslie 1988). The simultaneity between EUV flash and ZP burst
implies that the source region of the ZP burst is most possibly
very close to the small EUV flash with a temperature from several
decades of thousand K (such as at 304 \AA) up to $10^{7}$ K (193
\AA~ etc).

From the width and length of the EUV flash and the microwave
intensity of the ZP burst, we can obtain another estimation of the
emission brightness temperature: $1.84\times10^{11}$ K, which is
very close to the above result ($2.10\times10^{11}$ K) obtained
from the frequency bandwidth and the intensity of the ZP burst.
Both estimations of the brightness temperatures exceed 5 orders of
magnitude of the plasma thermal temperature, and this implies that
the microwave ZP burst is a coherent emission process, which can
amplify the initial fluctuations over several orders of magnitude
(Dulk 1985; Bastian, Benz, \& Gary 1998), and this coherent
process is possibly located in the small region around the above
EUV flash.

According to the recent classification of microwave ZPs (Tan et
al. 2014), the ZP burst here belongs to equidistant ZP, which is
possibly originated from Bernstein wave mechanism (Rosenberg \&
Tarnstein 1972; Zheleznyakov \& Zlotnik 1975; Maltseva \& Chernov,
1989). In this mechanism, the energetic electrons with
non-equilibrium distribution over velocities perpendicular to the
magnetic field are located in a small source, where plasma is
weakly and uniformly magnetized ($f_{pe}\gg f_{ce}$). These
electrons excite longitudinal electrostatic waves at the sum of
Bernstein mode frequency $sf_{ce}$ and the upper hybrid frequency
$f_{uh}$: $f=f_{uh}+sf_{ce}\approx f_{pe}+sf_{ce}$. Here, $f_{pe}$
is the electron plasma frequency, $f_{ce}$ is the electron
gyro-frequency, $s$ is the harmonic number. In such mechanism, all
zebra stripes are generated from a small compact source (this
point is compatible to the small EUV flash), and the emission is
coherent, which is much stronger than that of the gyrosynchrotron
radiation from the energetic electrons spiraling along magnetic
field lines in the flaring region. The frequency separation
between the adjacent zebra stripes is just the electron
gyro-frequency: $\bigtriangleup f=f_{ce}$, it is a constant. By
using the stripe frequency separation (90 MHz), we may directly
measure the magnetic field strength in the source region. The
value is about 32 Gs. From the emission frequency, the
corresponding plasma density can be estimated as $8.4\times10^{10}
cm^{-3} - 1.3\times10^{11}cm^{-3}$. These values are close to the
conditions of solar corona near the flaring regions. The Bernstein
wave mechanism requires large number of energetic electrons in a
small-size compact source region. The production of these
energetic electrons implies a fast particle acceleration taking
place in the source region, and their propagation will carry out
energy effectively from the source region which means a strong
energy release process.

Generally, the flaring energy is released from magnetic
reconnection processes, and the large proportion of energy is
first released into energetic non-thermal particles
(Tandberg-Hanssen \& Emslie 1988; Dere, Bartoe, \& Brueckner 1989;
Lin, Soon, \& Baliunas 2003; Saint-Hilaire, \& Benz 2005;
Aschwanden, Dennis, \& Benz 1998). These non-thermal particles can
produce two kinds of emission bursts: (1) microwave bursts, when
the non-thermal particles interact with the ambient plasmas, they
may produce strong microwave burst by form of coherent mechanism
which is always triggered by certain plasma instabilities. In this
work, the non-thermal electrons are sufficient to trigger
Bernstein waves, and the coupling between Bernstein wave and
Langmuir waves makes the microwave bursts forming the extremely
strong ZP burst. (2) EUV bursts, when the large number of
non-thermal electrons precipitate and impact to the ambient
plasma, they may deposit their energy and heat the plasma up to
more than several million Kelvin by collision mechanism, and
produce the strong EUV flashes. Both of the above two bursts make
up the very small and super strong ZP explosion.

Therefore the impulsive EUV flash associated with the super strong
ZP burst is a sudden explosion which can be regarded as a prompt
signature of explosive energy release in the above small region.
In fact, the flaring source region begins to brighten up from
several small FBPs on the two ribbons before the flare onset (e.g.
06:57:37 UT). It is reasonable to suppose that these FBPs are the
initial sites of magnetic reconnection and energy release. The ZP
burst is just the strongest one which produced great number of
energetic electrons and released powerful energy in a short time,
and triggered the coherent microwave burst.

\section{Conclusions and Discussions}

From the above observations and physical analysis, we may get the
following conclusions:

(1) A very short and super strong ZP burst observed simultaneously
by two microwave telescopes located in China and Czech Republic
just at the beginning of a solar M6.5 flare. This microwave ZP
burst is so strong that its emission intensity exceeds several
times of the underlying flaring continuum emission, and its
brightness temperature exceeds $10^{11}$ K. It is possibly a
coherent process triggered by the interaction between large number
of non-thermal electrons and plasma instability in a small compact
region. The non-thermal electrons are accelerated from a rapid
magnetic reconnection in the source region.

(2) During the same time of the ZP burst, an EUV flash takes place
with a sudden extra enhancement overlapped on the background
continuous brightening observed at EUV images by AIA/SDO. The
emission exceeds the saturation level at some EUV wavelengths. The
flash region is only a small short bright point in the flaring
region. Its existence indicates that there may be a rapid plasma
heating and fast energy release in the source region.

(3) The simultaneity of the strong microwave ZP burst and the
small EUV flash indicates that they may share a same small scale
source region where rapid particle acceleration, violent energy
release, and fast plasma heating take place simultaneously. It is
an unusual explosion occurring in a small region centered in the
flare source region just at the beginning of the solar flare.

The above results also indicate that a major solar flare may start
and develop from many small-scale bursting points, such as FBPs in
this work. The small EUV flash associated to the ZP burst produce
an extremely strong explosion in dimension of less than
2200$\times$5000 km and plasma temperature up to several $10^{6}$
K. The ZP burst is a strong coherent emission process which will
transport energy away rapidly from the source region in form of
energetic particles. And the small EUV flash reflects a fast
conversion from non-thermal particles to heat the coronal plasmas
to very hot in a small region and short period. The former may
imply a rapidly particle acceleration in a small region, while the
latter may imply a fast heating process in the solar atmosphere.

From letters of Drs. Bin Chen and Brian R. Dennis, they found that
there was an obvious enhancement at 12-25 keV and 25-50 keV of
RHESSI hard X-ray emission during the Zebra burst, and the
homochronous RHESSI image at 25-50 keV shows that a non-thermal
HXR footpoint source roughly coincides with the EUV footpoint
brightening. This fact may be another evidence supporting the
non-thermal origin of the ZP burst.

So far, as we have no high-resolution imaging observations at the
corresponding microwave frequencies, we have no direct way to get
the information of the source region of microwave ZP burst. When
the new generation solar radio heliograph (such as the Chinese
Spectral Radio Heliograph, CSRH, Yan et al. 2009) come into
service, we may have much more opportunity to obtain directly the
locations, geometrical structures, and magnetic fields in the
source region of the microwave ZP bursts.

\acknowledgments The authors would like to thank the referee for
helpful and valuable comments on this paper. Drs. Bin Chen and
Brian R. Dennis provide important results of RHESSI hard X-ray
emission after the acceptance of this paper. We also thank the
GOES, SDO, ORSC/Ond\v{r}ejov and SBRS/Huairou teams for providing
observation data. This work is supported by NSFC Grant 11273030,
11103044, 11103039, 11221063, 11373039, MOST Grant 2011CB811401,
the National Major Scientific Equipment R\&D Project ZDYZ2009-3,
and the Grant P209/12/00103 (GA CR). This work was also supported
by the Marie Curie PIRSES-GA-295272-RADIOSUN project.

\end{document}